# RADIOBIOLOGICAL IMPACT EVALUATION WITHIN MONTE-CARLO SHIELDING CALCULATIONS OF CANDU SPENT FUEL

*M. Sallah [1,2], C. A. Mărgeanu [2], N. Elbassiony [1], M. Mitwalli [1]* , *A. Elgarayhi [1]*

[1] Physics Department, Faculty of Science, Mansoura University, Mansoura 35516, Egypt
[2] Higher Institute of Engineering and Technology, New Damietta, Egypt
[3] Institute for Nuclear Research (RATEN ICN) Pitești, Romania

ABSTRACT

The radiobiological effect on the human health of CANDU spent fuel is assessed using Monte Carlo shielding estimates. The examination of spent fuel occurs after it has been discharged from the reactor. A specific cooling interval is considered, with the radiation dose rates that characterize the used fuel being of interest. Two kinds of fuel were studied in a CANDU standard fuel bundle with 37 fuel components: natural uranium (NU) fuel and slightly enriched uranium (SEU) fuel. The fuel burnup was simulated using the ORIGEN-S algorithm, and the photon sources describing the wasted fuel were retrieved. A generic stainless steel shipping cask type B was used for spent fuel transfer, and radiation doses at the cask wall and in the air up to 8 m away from the shipping cask were computed using the Monte Carlo MORSE-SGC algorithm. To ensure nuclear safety and radiation protection, spent fuel must be maintained in temporary wet cooling storage for six months. The projected dosage rates were modest, allowing for the safe handling of the used fuel shipping cask. The corresponding dosages on human body organs for the two considered spent fuels were estimated without and with shielding. Due to the varied sensitivity and reaction of organs/tissue, the effective dosage was evaluated for the human body by applying a tissue-weighting factor; these weighting factors are not equal, and functional coefficients specified by ICRP are used. The equivalent doses calculation modeling findings for the present study underlined the complete effectiveness of the applied shielding and attained the acceptable dose level.

## 1. Introduction

After the spent nuclear fuel is discharged from the reactor, a series of challenges for nuclear safety should be faced, as follows: high radioactivity due to fission products disintegration; the presence of plutonium and unconsumed uranium; release heat of the hot fuel, criticality control, reduction of the waste amounts from reactor operation, protection of personnel and public living near the nuclear facility, environmental impact, impact of radioactive waste on future generations [1-5]. In terms of nuclear safety and radiation protection, spent nuclear fuel should be kept in temporary wet storage inside the reactor pool for cooling (spent fuel decreasing) and reducing the radioactivity at levels allowing its safe manipulation and further storage in dedicated storage facilities. The spent fuel mandatory cooling period for CANDU reactors is for at least six months [3, 5-7].

Radioactivity is the macroscopic expression of nuclear decay, which is the physical process occurring spontaneously in a stochastic manner when atomic nuclei of an isotope undergo internal transformations to achieve a more stable energy state. The decay process is accompanied by the emission of nuclear particles or photons carrying the energy in excess. Thus, the estimation of radioactivity aims to identify and quantify radioactive isotopes since nuclear radiation may occur in various types, abundances, and energies, which are characteristic of each radionuclide. Therefore, it is important to calculate the radiobiological impact of any radiation source [8-13].

Radiation protection during the spent nuclear fuel management is very important in terms of human health and environmental effects, considering the activities performed by the personnel during the spent fuel discharged from the reactor core, cooling inside the reactor pool, transport, and storage in dedicated facilities. The radiation dose rates compliance with the internationally recommended limits is assured based on the corresponding shielding calculations [8, 13]. The paper's main objective was to evaluate the radiological impact on personnel following potential exposure to radiations emitted by the CANDU spent fuel discharged from the reactor. It cooled down up to 5 years in intermediate wet storage (inside the reactor's pool). Effective doses on human body organs were calculated based on the radiation dose rates estimated through previously performed shielding calculations for a CANDU standard fuel bundle containing natural uranium (NU) or slightly enriched uranium (SEU) fuel, irradiated in CANDU reactors specific conditions [3]. The spent nuclear fuel is cooled down inside the reactor's pool, with concrete walls stainless steel reinforced, at proper distances to avoid criticality issues, underwater that functions both as shielding material and coolant. For each step of the shielding calculations, the radiobiological impact and effective dose rates on the human body organs are estimated both to validate the efficiency of the shielding and to emphasize the mandatory need for radiation protection to preserve human health [3].

## 2. The theoretical model for shielding calculations

CANDU standard fuel bundle has a cylindrical shape. It comprises 37 fuel elements (tubes of zircaloy, loaded with fuel pellets) arranged on three concentric rings (of 6, 12, and 18 elements, respectively) and one central element, as presented in Figure 1. The fuel element made of zircaloy is 50 cm long, has an external diameter of 13 mm, and can accommodate a column of about 30 sintered fuel pellets; the sheath thickness is 0.4 mm [3].

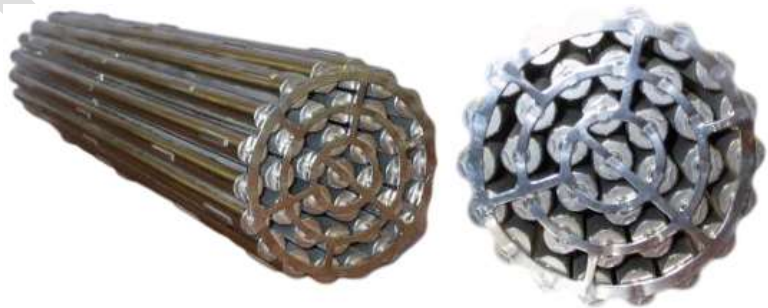

*Fig. 1*. *CANDU standard fuel bundle view [3].*

The proposed analyses were performed for two fuels, namely natural uranium (NU) fuel and slightly enriched uranium (SEU) fuel with 1.1 wt % enrichment in $^{235}$U; NU-37 and SEU-37 were used to identify the two considered CANDU fuels [3].

A geometrical model was developed to perform the shielding calculations considering the source of radiation (one fuel bundle) and the shielding (a generic shipping cask type B). The fuel bundle was irradiated in CANDU reactor conditions. After the "useful life" was reached, the spent fuel was discharged from the reactor, and a cooling period of up to 5 years was considered (within the reactor's spent fuel pool) [3]. The shielding calculations were performed using a generic shipping cask type B with stainless steel walls, which is recommended to transport fissile materials. The geometrical and material data were taken from a shipping cask prototype designed, manufactured, and tested at RATEN ICN Pitești [3, 14-16]. The cask model developed for the shielding calculations consists of a source region and a container (cask) region. The computing efficiency was improved considering the assumption of cask symmetry to its mid-plane; this assumption was used to implement the automated biasing procedure in the Monte Carlo analysis [14, 16, 17]. The geometrical model for the source of radiation was developed considering the fuel bundle composed of three straight, concentrically positioned cylinders separated by thin air gaps, as follows: a central cylinder for the main element and the inner ring of 6 fuel elements, the second cylinder for the intermediate round of 12 fuel elements, and the third cylinder for the outer ring of 18 fuel elements. In these cylinders, a homogenous mixture of fuel, clad and structural materials was considered preserving the volume. A simplified geometrical model was developed for the cask, composed of straight, concentrically positioned cylinders of shielding materials, with a central cavity for the source of radiation. Figure 2 illustrates the source of radiation and container geometrical models successfully used for shielding calculations in [1-5, 14, 16].

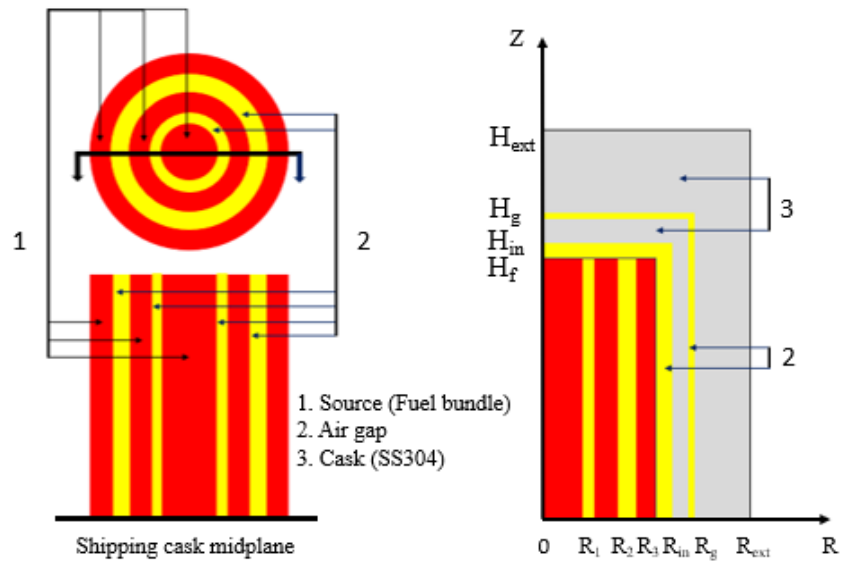

***Fig.2.*** *Geometrical configurations for the radiation source (left) and the cask (right),* [1-5, 14].
*$H_f$ – fuel cylinders height; $R_1$, $R_2$, $R_3$ – fuel cylinders radii; $R_{ext}$ – cask radius; $H_{ext}$ – cask height;*
*$R_{in}$ – cavity radius; $H_{in}$ – cavity height; $R_g$ – cask air gap radius; $H_g$ – cask air gap height*

### 3. Shielding calculations results

CANDU nuclear fuel irradiation has been simulated using the ORIGEN-S burnup code, included in the SCALE 5.1 programs package [18-20], considering specific fuel burnup powers of 34.644 kW/kg HE for NU, and 43.24 kW/kg HE for SEU, respectively (HE – heavy element). The ″useful life″ for the considered fuels was reached after 231 days of irradiation for NU (up to 8 MWd/kgHE burnup), and 278 days of irradiation for SEU (up to 12 MWd/kgHE), respectively. The photon radiation sources estimated during the cooling period were further used as input data for the shielding calculations [3].

The shielding calculations have been performed utilizing Monte Carlo MORSE-SGC code, also included in SCALE 5.1 programs package [19], using the (27n-18g) coupled nuclear data library (27 neutron and 18 gamma energy groups) and the ANSI standard flux-to-dose conversion factors (dose rates will be in rem/h). For the Monte Carlo simulation, 1000 bunches of 3000 particles have been generated [3]. The radial photon dose rates characterizing the spent fuel after 0.5-year, 1 year and 5 years of cooling were estimated at the cask wall and in air at 0.5 m, 1 m and 2 m distance from the cask, respectively [3, 18, 20-22].

NU and SEU fuel compositions are presented in Table 1. Each fuel establishes first the weight of the components in the fuel, then the isotopic consequences of $^{234}$U, $^{235}$U, $^{236}$U, $^{238}$U, and $O_2$ in the fuel, and finally, the initial inventory of materials. The fuel bundle clad is made of Zircaloy (Zy), containing Zirconium, alloying elements (Tin, Iron, Chromium, Nickel), and impurities (Carbon, Oxygen).

**Table 1.** Initial inventory in NU-37 and SEU-37 fuels [3].

| Fuel project | Enrichment in $^{235}$U | $^{234}$U | $^{235}$U | $^{236}$U | $^{238}$U | $O_2$ | Zy | $UO_2$ |
|---|---|---|---|---|---|---|---|---|
| | (wt%) | (kg) | | | | | | |
| *NU-37* | 0.71 | 0 | 0.138 | 0 | 19.357 | 2.62 | 2.144 | 22.115 |
| *SEU-37* | 1.1 | 0.015 | 0.175 | 0.064 | 19.151 | 2.608 | 2.144 | 22.013 |

The total fissile content was higher in the SEU-37 fuel bundle (~ 54 g/assembly) than in the NU-37 fuel bundle (~ 51 g/assembly). A relative difference of 6% is obtained. The fissile inventory was reduced as the fuel irradiation increased due to the higher consumption of $^{235}$U than the accumulation of $^{239}$Pu. Fig. 3 illustrates the radiation emission rates evolution with the energy of the photons obtained for NU-37 and SEU-37 spent fuels considering four cooling times (0.5 years, 1 year, 3 years, and 5 years, respectively) [3]. As the cooling time became longer, the radiation emission rates of spent fuel evolution with the cooling time showed a decreasing trend. After 5 years of cooling the photon emission rates became 70% - 97% lower than those obtained after six months. Similar profiles for both spent fuels can be noticed; however, the radiation emission rates characterizing SEU-37 spent fuel were about 30% higher than the NU-37 spent fuel ones [3].

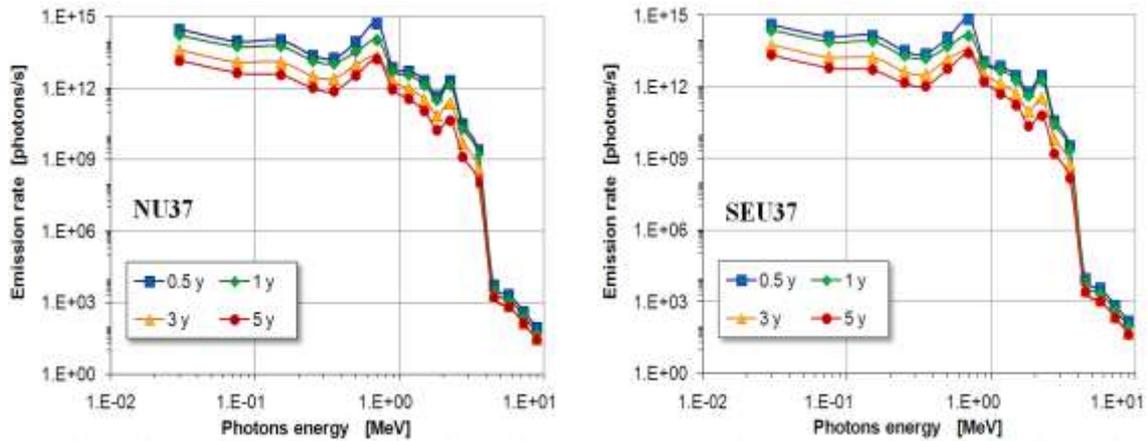
***Fig.3.** Radiation source profile for NU-37 (left) and SEU-37 (right) spent fuels [3]*

As mentioned before, four main measuring points for photon dose rates were proposed in [3], namely: the shipping cask wall and 0.5, 1, and 2 m (in air) distance from the cask. The minimum (cask wall) and the maximum (2 m distance from the cask) measuring points were chosen to check the safety of spent fuel transport in terms of compliance with the international recommended limits [8, 13, 23, 24]. Photon dose rates estimated for NU-37 and SEU-37 fuel bundle projects, considering before mentioned cooling times and measuring points, have been calculated and are presented in Figure 4.

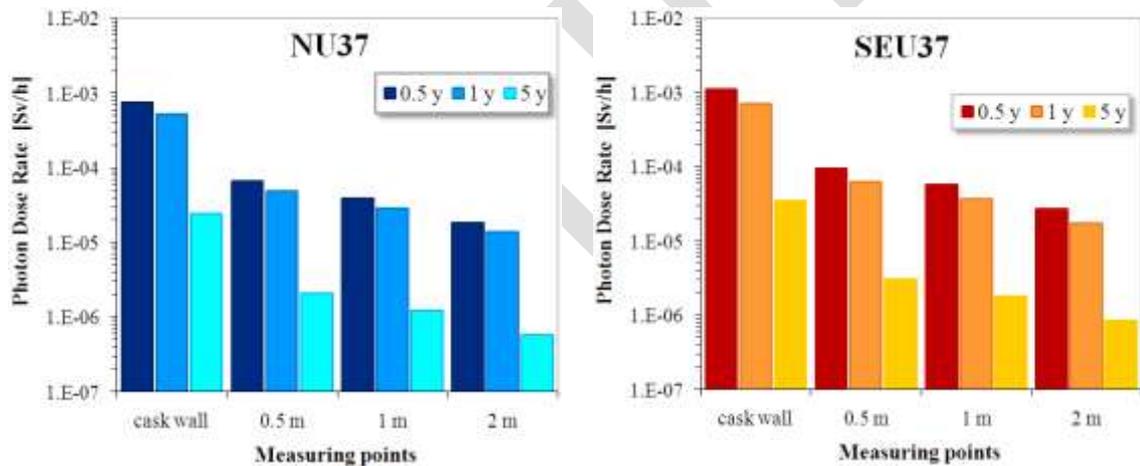
***Fig.4.** Photon dose rates with cooling time and measuring points for NU and SEU spent fuels [3].*

Estimated photon dose rates were lower as the cooling time became longer and significantly decreased with increasing distance from the shipping cask wall. After 6 months of cooling, the photon dose rates to the cask wall were 0.76 mSv/h for NU-37, and 1.3 mSv/h for SEU-37 spent fuels [3]. After 5 years of cooling, the photon dose rates to the cask wall were 0.024 mSv/h for NU-37, and 0.035 mSv/h for SEU-37 spent fuels [3, 20, 24, 25]. These values are drastically reduced as the distance from the external cask surface increased, such as at 2 m distance from the cask, the photon dose rates became 0.018 mSv/h (after 6 months of cooling) and 0.006 mSv/h (after 5 years of cooling) for NU-37 spent fuel, and 0.027 mSv/h (after 6 months of cooling) and 0.008 mSv/h (after 5 years of cooling) for SEU-37 spent fuel, respectively [3].

The radiation dose rates obtained for SEU-37 spent fuel were 30% higher than the NU-37 ones. However, for both NU-37 and SEU-37 spent fuels, safe values were obtained, the estimated photon dose rates being lower than the internationally agreed limits for safe transport of radioactive material using the shipping cask type B. These internationally agreed transport regulations for radioactive material establish that the radiation level under normal conditions of transportation shall not exceed 2 mSv/h to the cask wall, and 0.1 mSv/h at 2 m from the external surface of the conveyance [2, 3, 8, 13, 23].

## 4. Distribution pattern of effective dose to body's Organ/Tissue

The effective dose is considered for the whole body by applying a tissue-weighting factor (WT) of 1; calculating the dose on any specific tissue inside the body is a fraction of 1. The sensitivity of hard/soft tissue varies from one organ to another depending on the absorbed rate; hence the equivalent dose will vary. Therefore, the tissue weighting factors are not equal and, according to ICRP recommendations [12, 13], are divided into three categories: low risk (0.01), medium risk (0.05), and high risk (0.12). The organs distribution pattern of the effective dose ($D_{organ}$) for the body organ shall be calculated from

$$D_{organ} = D_{eqv} F_{WT} \quad (mSv/h)$$

where: $D_{eqv}$ is the equivalent dose emitted by the spent fuel with/without shielding. The tissue-weighting factor (FWT) differs from one organ/tissue to another [13], and the utilizing coefficients are shown in table 2.

**Table 2.** The conversion coefficient is used for the calculation distribution pattern of dose.

| Organ/tissue | $F_{WT}$ |
|---|---|
| Gonads | 0.20 |
| Stomach, colon, lung, and red bone marrow | 0.12 |
| Liver urinary bladder, esophagus, thyroid gland, and breast | 0.05 |
| Other body organs | 0.05 |
| Skin and bone surface | 0.01 |

The calculations of the equivalent dose for stomach, colon, lung, and red bone marrow corresponding to NU-37 and SEU-37 without shielding i.e., before utilizing the shielding cask are shown in Table 3. The mentioned equivalent dose of organs is classified into three scheduled calculations depending on the duration of cooling time (0.5, 1, and 5 years). For 6 months of cooling, the effective dose for NU-37 and SEU-37 spent fuels considering different measuring points (from 0.5 m to 8 m) with averages of 202.91 and 267.85 mSv/h, respectively. For one year of cooling, the corresponding doses of NU and SEU spent fuels have averages of 52.99 and 74.80 mSv/h, respectively. The average doses for 5 years of cooling decreased NU and SEU, being 7.16 and 11.13 mSv/h, respectively. The equivalent doses without shielding indicate very high values compared to the world limitation and permissible dose recommended from ICRP and IAEA [8, 13, 25]. Although the dose levels drop according to the inverse square law, and the half-life of radioactive isotopes is lengthy, thus the research depended on overcoming this issue by using shielding calculations and demonstrating this using radiological impact, as shown in Table 4.

**Table 3.** The equivalent dose of gonads regarding exposure to each NU37 and SEU37 (*mSv/h*) without shielding.

| D/Y | 0.5 y | | 1 y | | 5 y | |
|---|---|---|---|---|---|---|
| | NU37 | SEU37 | NU37 | SEU37 | NU37 | SEU37 |
| 0.5 m | 780.4 | 1036.4 | 204.4 | 288.6 | 27.64 | 42.88 |
| 1 m | 426.8 | 564.8 | 111.64 | 157.36 | 15.08 | 23.42 |
| 2 m | 178.58 | 234.4 | 46.52 | 65.62 | 6.29 | 9.78 |
| 3m | 95.04 | 123.98 | 24.64 | 34.88 | 3.35 | 5.19 |
| 4m | 58.4 | 75.6 | 15.08 | 21.38 | 2.05 | 3.18 |
| 5m | 39.32 | 50.64 | 10.15 | 14.38 | 1.38 | 2.14 |
| 6m | 28.22 | 36.08 | 7.28 | 10.24 | 0.98 | 1.53 |
| 8m | 16.52 | 20.92 | 4.24 | 5.98 | 0.58 | 0.89 |

Given the shielding calculations results presented in section 3, the equivalent doses using the shielding cask for different measuring points are shown in Table 4 for the gonads. Even for short-term cooling, the shielding gives excellent results as long as the equivalent dose decreases and becomes permissible at a distance of 50 cm from the cask wall. Noteworthy, the equivalent dose becomes negligible for more than 2 m from the cask wall. It gives significant results, particularly for nuclear and radioactive materials [5 and 6]. The results shown in Fig. (5, a-d), confirm that the equivalent doses are under the permissible dose rate limit starting from cooling times longer than 6 months at 0.5 m distance from the cask wall for both NU-37 and SEU-37, for 3 m distance from the cask wall the equivalent dose being approximately zero.

**Table 4.** The equivalent dose of gonads regarding exposure to each NU37 and SEU37 (*mSv/h*) with shielding.

| D/Y | 0.5 y | | 1 y | | 5 y | |
|---|---|---|---|---|---|---|
| | NU37 | SEU37 | NU37 | SEU37 | NU37 | SEU37 |
| Cask wall | 0.15226 | 0.221 | 0.10372 | 0.14326 | 0.00481 | 0.00707 |
| 0.5 m | 0.01329 | 0.01932 | 0.00977 | 0.01262 | 0.00042 | 0.00061 |
| 1 m | 0.00788 | 0.0114 | 0.00579 | 0.00748 | 0.00025 | 0.00036 |
| 2 m | 0.00372 | 0.00537 | 0.00275 | 0.00352 | 0.00012 | 0.00017 |
| After 2 m | 0 | 0 | 0 | 0 | 0 | 0 |

For investigated the efficiency of the shielding all conversions coefficient of Table 2. has used for the calculation distribution pattern of effective dose and presented graphically as shown in fig. the calculated infographics assure that the effective doses of organs are transferred to a low level and become permissible for safe manipulation. Moreover, the result presented product (cask) achieves safety and security standards for professionals and the public in the shortest distances from the cask wall and in the shortest possible time for cooling the separated fuel.

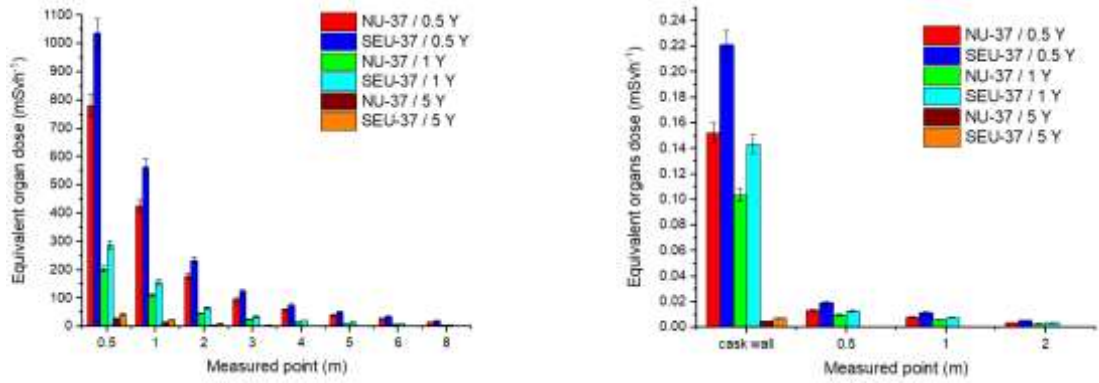

***Fig 5.a:*** *Dose of gonads (mSv/h) as without shielding (left) and with shielding (right)*

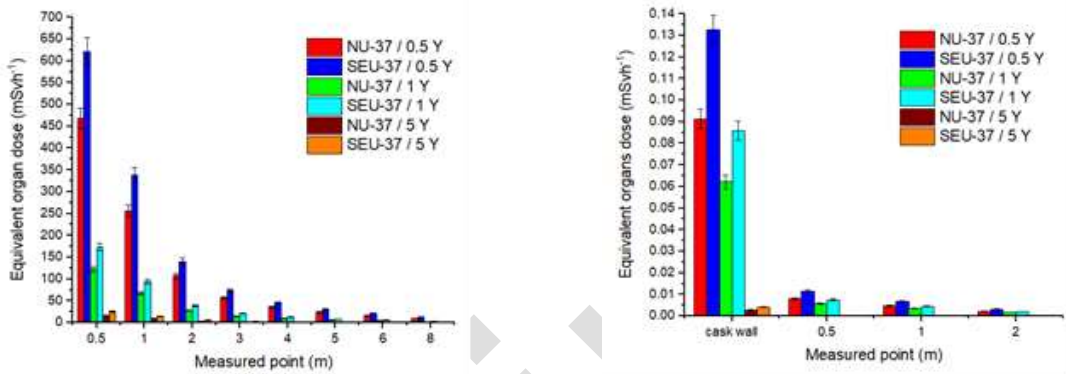

***Fig 5.b:*** *Dose of stomach categories (mSv/h) as without shielding (left) and with shielding (right)*

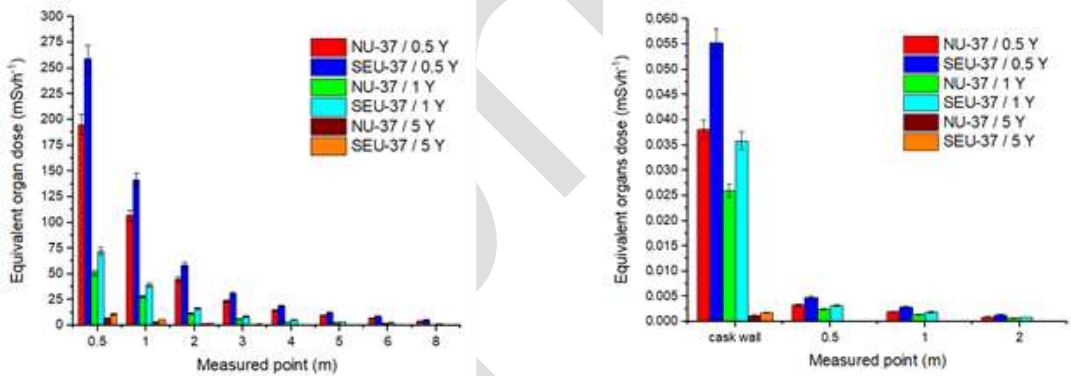

***Fig 5.c:*** *Dose of liver categories (mSv/h) as without shielding (left) and with shielding (right)*

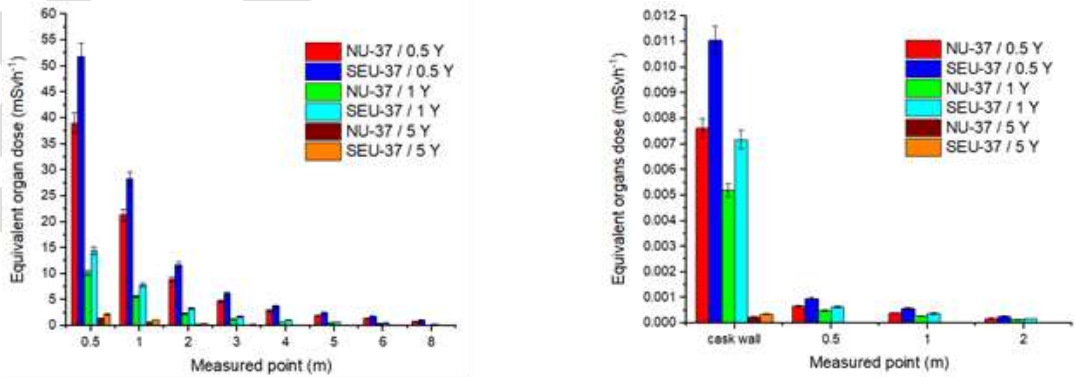

***Fig 5.d:*** *Dose of skin categories (mSv/h) as without shielding (left) and with shielding (right)*

The research data has furnished for human safety and organ doses in terms of excess lifetime cancer risk (ELCR) are calculated and presented in Fig. 6 for NU-37 and SEU-37 without and with shielding. ELCR denotes the excess probability of developing cancer during the lifetime due to natural gamma radiation exposure to the resident population by radionuclides content of the sample and can be calculated from the equation [5 and 7]

$$\text{ELCR} = \text{ED (Sv/y)} \times \text{DL (70y)} \times \text{RF (0.05 Sv}^{-1}\text{)}$$

ED is the annual effective dose (Sv/year), DL duration of life (averaged as 70 years old). RF denotes the risk factor for 1 Sievert absorbed dose, and it is calculated by the International Commission of Radiological Protection (ICRP) with the value $0.05$ $Sv^{-1}$ [12, 13].

ICRP calculates the permissible ELCR to have an average $3.65 \pm 1.85 \times 10^{-3}$. Hence, Fig. 6 represents the evolution of ELCR for both NU-37 and SEU-37 without and with shielding. Even for cooling of 5 years for the spent fuel without shielding, ELCR is still very high even at 8 m far from the cask wall. Consequently, the designed shielding of the CANDU spent fuel allows estimating an excellent radiobiological safety for the working personnel. The corresponding descriptive statics of the ELCR is shown in Fig. 7 to assure the safety and efficiency of working with the shielding without worries.

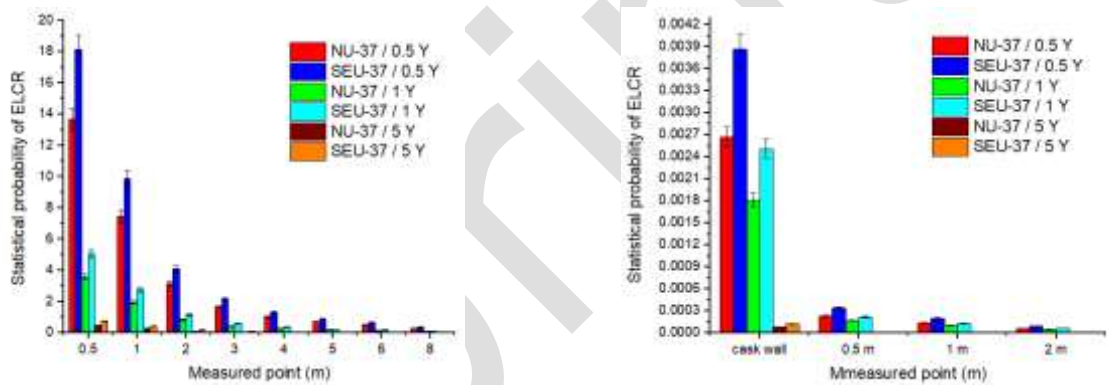

***Fig. 6.*** *Comparison between ELCR at different distances for both NU-37 and SEU-37 without shielding (left) and after the shielding (right) (mSv/h)*

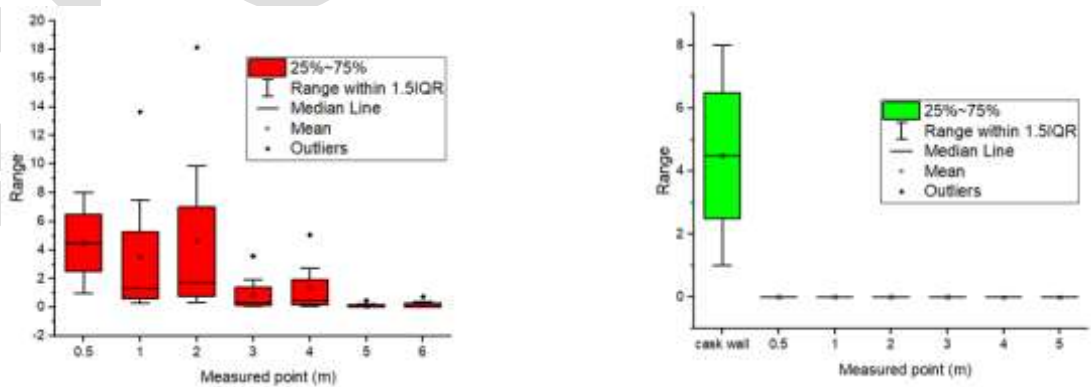

***Fig. 7.*** *ELCR descriptive statics without shielding (left) and with shieling (right)*

**Conclusions**

The present study's main goal was to evaluate the radiobiological impact of CANDU spent fuel on the human body organs based on previous performed Monte-Carlo shielding calculations. The radiation source was a single CANDU standard fuel bundle with 37 fuel elements containing natural uranium fuel (NU) or slightly enriched uranium (SEU) with 1.1 wt% enrichment in $^{235}$U, respectively. The considered fuels were irradiated according to CANDU reactors specific to 8 MWd/kg HE (NU fuel) and 12 MWd/kg HE (SEU fuel), respectively.

The spent fuel was discharged from the reactor and cooled for up to 5 years to allow the fission products to disintegrate and spent fuel radioactivity and heating reduced. The photon dose rates were estimated using MORSE-SGC Monte Carlo code both to the shipping cask wall and in the air at different distances from the external cask surface to assure radiation protection of personnel during spent fuel transport to storage facilities. Estimated photon dose rates were lower as the cooling time increased and were significantly reduced as the distance to the shipping cask wall became greater. SEU-37 spent fuel radiation dose rates were ~ 30% higher than those estimated for the NU-37 spent fuel. Both considered fuel compositions had lower values than the internationally agreed limits for safe transport of radioactive material using the shipping cask type B.

Based on the shielding calculations conducted with protection and safety, the distribution pattern of the organs dose was calculated and reflected typical safety and radiation protection for human health. The overall effective dose values of shielding application compared with the recommended threshold are emphasized, which enabled us to estimate the impact of the radiobiological hazard, which determined the safe distance limit of 2 m, where the dose became negligible, principally feature of short cooling time as well as the calculation property of ELCR indicated a safe side for long-term monitoring, which helps to perform good protection protocol for workers from cancer risk. Our future work will consider the shielding calculations and radiobiological impact for different types of spent fuel and other sources of radiation and radioactive materials.